\documentclass[11pt, a4paper]{article}
\pdfoutput=1
\usepackage{jcappub}
\usepackage{geometry} 
\usepackage{url}
\usepackage{amsmath}
\usepackage{amssymb}
\usepackage{graphicx}
\usepackage{hyperref}
\usepackage{siunitx}
\usepackage{upgreek}
\usepackage{xcolor}

\newcommand{\subrm}[1]{_\mathrm{#1}}
\newcommand{\submax}{_\mathrm{max}}

\newcommand{\subeff}{_\mathrm{eff}}
\newcommand{\diff}{\mathrm d}
\newcommand{\up}{\mathrm}
\newcommand{\mr}{\mathrm}

\title{Statistical sensitivity on right-handed currents in presence of eV scale sterile neutrinos with KATRIN}

\author[a, d]{Nicholas M. N. Steinbrink, }
\author[b]{Ferenc Gl\"uck, }
\author[c]{Florian Heizmann, }
\author[c]{Marco Kleesiek, }
\author[b]{Kathrin Valerius, }
\author[a]{Christian Weinheimer, }
\author[d]{and Steen Hannestad}

\affiliation[a]{Institute for Nuclear Physics, University of M\"unster\\ Wilhelm Klemm-Str. 9, 41849 M\"unster, Germany }
\affiliation[b]{Institute for Nuclear Physics, Karlsruhe Institute of Technology \\ P.O. Box 3640, 76021 Karlsruhe, Germany}
\affiliation[c]{Institute of Experimental Nuclear Physics, Karlsruhe Institute of Technology\\ P.O. Box 3640, 76021 Karlsruhe, Germany }
\affiliation[d]{Department of Physics and Astronomy, Aarhus University \\ Ny Munkegade 120, 8000 Aarhus C, Denmark }

\emailAdd{n.steinbrink@uni-muenster.de}
\emailAdd{ferenc.glueck@kit.edu}
\emailAdd{florian.heizmann@kit.edu}
\emailAdd{marco.kleesiek@kit.edu}
\emailAdd{kathrin.valerius@kit.edu}
\emailAdd{weinheimer@uni-muenster.de}
\emailAdd{steen@phys.au.dk}

\abstract{
	The KATRIN experiment aims to determine the absolute neutrino mass by measuring the endpoint region of the tritium $\upbeta$-spectrum. As a large-scale experiment with a sharp energy resolution, high source luminosity and low background it may also be capable of testing certain theories of neutrino interactions beyond the standard model (SM). An example of a non-SM interaction are right-handed currents mediated by right-handed W bosons in the left-right symmetric model (LRSM). In this extension of the SM, an additional SU(2)$\subrm R$ symmetry in the high-energy limit is introduced, which naturally includes sterile neutrinos and predicts the seesaw mechanism. In tritium $\upbeta$ decay, this leads to an additional term from interference between left- and right-handed interactions, which enhances or suppresses certain regions near the endpoint of the beta spectrum. In this work, the sensitivity of KATRIN to right-handed currents is estimated for the scenario of a light sterile neutrino with a mass of some eV. This analysis has been performed with a Bayesian analysis using Markov Chain Monte Carlo (MCMC). The simulations show that, in principle, KATRIN will be able to set sterile neutrino mass-dependent limits on the interference strength. The sensitivity is significantly increased if the $Q$ value of the $\upbeta$ decay can be sufficiently constrained. However, the sensitivity is not high enough to improve current upper limits from right-handed W boson searches at the LHC.
}


\arxivnumber{1234.5678}

\begin{document}

\maketitle

\flushbottom

\section{Introduction}

The KArlsruhe TRItium Neutrino (KATRIN) \cite{KATRINDesignReport} is the most sensitive next-generation neutrino mass experiment currently under commission. It was designed to probe the absolute scale of the neutrino mass with a targeted sensitivity of \SI{0.2}{eV} at 90 \% confidence level by measuring the endpoint of the tritium $\upbeta$-decay spectrum. As a large-scale $\upbeta$-decay experiment with the most precise integral energy spectrometer to date with an energy resolution of up to \SI{0.93}{eV} at the $\upbeta$ endpoint \cite{Groh2015} and the so far most luminous tritium source with an activity of about \SI{e11}{Bq} \cite{Babutzka2012}, KATRIN also has some capability of testing theories beyond the standard model.

The sensitivity of KATRIN to sterile neutrinos has been addressed by some publications in the past regarding the eV \cite{Formaggio2011, Riis2011, Esmaili2012} and the keV scale \cite{Mertens2015, Mertens2015a}. Several experimental anomalies, for instance from the reactor neutrino anomaly \cite{Mention2011}, the calibration of solar neutrino experiments GALLEX and SAGE  \cite{GALLEX1995, SAGE2006, SAGE2009, Kaether2010} and the short baseline accelerator neutrino oscillation experiments  LSND \cite{LSND1998} and MiniBooNE \cite{MiniBooNE2007}, might be explained by a sterile neutrino on the mass scale of a few eV \cite{Abazajian2012}. However, the hypothesis is difficult to reconcile with cosmology. Measurements of primal abundance of deuterium at the time of the Big Bang Nucleosynthesis (BBN) \cite{Cooke2014, Cyburt2016} and Cosmic Microwave Background (CMB) measurements with Planck \cite{Planck2016} suggest that the effective number of neutrino degrees of freedom in the universe is inconsistent with a fourth neutrino with a mass of only a few eV. The tension can in principle be solved, e.g., by assuming certain non-standard interactions \cite{Hannestad2014, Archidiacono2015, Archidiacono2016}. On the other hand, the hypothesis has further been challenged by IceCube which found no evidence for light sterile neutrinos \cite{IceCube2016}. The question might finally be solved by KATRIN which will be able to exclude nearly the complete parameter space from the reactor neutrino anomaly \cite{Formaggio2011, Riis2011, Esmaili2012}. Additionally, dedicated experimental efforts such as SOX \cite{SOX2016}, STEREO \cite{Stereo2015} and DANSS \cite{DANSS2016} are currently under way to test the sterile neutrino hypothesis in short-baseline oscillations. Analyses of data from the KATRIN predecessor experiments in Mainz \cite{Kraus2013} and Troitsk \cite{Belesev2013} could already exclude a small region of the parameter space.

Another interesting field where KATRIN might set new limits is represented by neutrino interactions beyond the standard model. Especially the sensitivity to weak non-$V\!-\!A$ contributions, e.g. right-handed currents, has been studied in several publications \cite{Stephenson2000, Severijns2006, Bonn2011, Riis2011a, Barry2014}. Aside from a model-independent study it is also worthwhile to investigate particular models in which right-handed currents arise naturally. 

The left-right symmetric model (LRSM) \cite{Pati1974, Mohapatra1975, Senjanovic1975} as such a general framework is a simple extension of the standard model, which adds an additional SU(2)$\subrm R$ symmetry acting only on right-handed fermion fields, analogous to the left-handed SU(2)$\subrm L$ of the standard model, thus restoring parity on high energy scales. The SU(2)$\subrm R$ is mediated by right-handed W and Z bosons which have masses at least in the TeV scale \cite{Maiezza2010}. Current experimental limits of the mass of the right-handed W from the LHC approximately give $m_{W_R} \gtrsim \SI{3}{TeV} $ \cite{CMS2014, ATLAS2012}. Depending on the parameters, the LRSM can lead naturally both to the seesaw mechanism (type I and II)  \cite{Mohapatra1980, Tello2011, Parida2013, Nemevsek2013, Barry2013} and sterile neutrinos on observable mass scales \cite{Bezrukov2010}. By means of the right-handed weak bosons, right-handed currents are an essential constituent of the LRSM \cite{Barry2014}, however strongly suppressed due to the high masses of the right-handed bosons. 

In this paper, the statistical sensitivity of KATRIN will be determined for a combined scenario of right-handed currents with light sterile neutrinos on the eV scale. On this scale, there are in principle no modifications of the hardware and the data acquisition required, which means that the analysis can be performed just with the data of the primary neutrino mass measurement runs. The scenario is motivated by the LRSM but shall be addressed in a way which is as model-independent as possible. Nevertheless, the results will be discussed with respect to the parameter space of the LRSM which is not yet experimentally excluded.

\section{Theory}

In the following, the tritium $\upbeta$-decay spectrum, as measured in KATRIN, will be described and the modifications with respect to the SM to include right-handed currents and eV scale sterile neutrinos, based on the LRSM, will be described. The kinematics of the modified $\upbeta$-spectrum will be discussed thereafter.

\subsection{Tritium $\upbeta$-decay}

The tritium $\upbeta$-decay spectrum with left-handed currents only is given in natural units (e.g., refs.~\cite{Otten2008, Drexlin2013}) as

\begin{align}
	\begin{split}
	w(E)  = \ & \sum_{i,j} |U_{\mathrm{e}i}|^2 \cdot P_j \cdot w_{ij}(E) \\
    \label{eq:beta}
    = \ & N \frac{G_\up F^2}{2 \pi^3} \cos^2(\theta_C) |M|^2 \ 
	F (E, Z') \cdot p \cdot (E+m\subrm e) \ \cdot \\
	& \sum \limits_{i,j} |U_{\mathrm{e}i}|^2 \cdot P_j \cdot (E_0 - V_j - E) \cdot \sqrt{(E_0 - V_j - E)^2 - m_i^2} \ ,
	\end{split}
\end{align}

where $E$ is the kinetic electron energy, $m_i$ the mass of the $i$-th neutrino mass eigenstate, $E_0-V_j$ the spectral endpoint, i.e., the maximum kinetic energy in case of $m_i= 0$, $|U_{\mathrm{e}i}|^2$ the mixing matrix element between the electron neutrino $\upnu \subrm e$ and the $i$-th mass eigenstate, $\theta_C$ the Cabbibo angle, $N$ the number of tritium atoms, $G_F$ the Fermi constant, $M$ the nuclear matrix element, $F (E, Z')$ the Fermi function with the charge of the daughter ion $Z^{'}$ and $p$ the electron momentum. If the source consists of gaseous molecular tritium, as in the KATRIN experiment \cite{Babutzka2012}, the spectrum needs to be summed over the electronic and rotational-vibrational final states of the daughter molecules as indicated in the first line of eq. \eqref{eq:beta}. That means that the final spectrum is a superposition of spectra $w_j(E)$ with different final states, where $P_j$ is the probability to decay to a state with excitation energy $V_j$ \cite{Saenz2000, Doss2006, Doss2008}. 

If there is no contribution by sterile neutrinos, only three mass eigenstates appear in eq. \eqref{eq:beta}. The LRSM assumes three additional sterile neutrinos \cite{Barry2013}. While in the most simple models the sterile neutrino masses are beyond the \si{TeV} scale to maintain the seesaw mechanism, it is also possible to have at least one light sterile neutrino \cite{Borah2016}. As we study the scenario of eV scale sterile neutrinos in conjunction with right-handed currents, we include only one additional mass state $ m_4 \sim \mathcal O(\si{eV})$. Since the light mass eigenstates 1, 2, 3 are not distinguishable by KATRIN \cite{Drexlin2013}, we can define a \emph{light neutrino mass} or  \emph{electron neutrino mass} as $m_l^2 \equiv \sum_{i=1}^3  |U_{\mathrm{e}i}|^2 m_i^2$. The \emph{heavy neutrino mass} or \emph{sterile neutrino mass} is then defined as  $m_h \equiv m_4$   and the \emph{active to sterile mixing angle} $\theta$ as $\sin^2\theta \equiv |U_{\mathrm{e}4}|^2$.

The modified $\upbeta$-spectrum with included right-handed currents based on a left-right symmetry has been derived in \cite{Barry2014} for a general case and been adopted for the special case of one sterile neutrino with keV mass. We use the same result of the derivation and apply a similar strategy for our scenario. We approximate the three light mass states by a single state $m_l$, as defined above, and take only one sterile state $m_h$ into account. Furthermore we ignore the possibility of CP-violating phases in the neutrino mixing matrix. The mixing matrix (eq.~2.2 in \cite{Barry2014}) then becomes a plain $2 \times 2$ rotation matrix. Additionally, since were are measuring on the eV scale, we take into the account the contribution of right-handed lepton vertices on the light neutrino, which has been neglected in  \cite{Barry2014} (eq.~3.10). The $\upbeta$-spectrum then takes the form

\begin{equation}
\begin{split}
w(E)  =  \sum_j P_j \cdot \bigg[ \ & w_{h_j}(E)  \cdot 
(a_\up{LL} \ \sin^2\theta + a_\up{RR} \ \cos^2 \theta) \\
 + \ & w_{l_j}(E)  \cdot  
(a_\up{LL} \ \cos^2\theta + a_\up{RR} \ \sin^2 \theta) \\
+ \ &  w_{h_j}(E)  \cdot  
\frac{m_h}{E_0-V_j-E} \cdot \frac{m_\up e}{m_\up e+E} \ a_\up{LR} \ \cos \theta \sin \theta \\
 - \ &w_{l_j}(E)  \cdot 
\frac{m_l}{E_0-V_j-E}  \cdot \frac{m_\up e}{m_\up e+E} \ a_\up{LR} \ \cos \theta \sin \theta  \ \bigg] \ ,	
\label{eq:beta-rh-theo}
\end{split}
\end{equation}
where the abbreviations $w_{l_j}(E)$ and $w_{h_j}(E)$ denote $j$-th final state component of the $\upbeta$-spectrum \eqref{eq:beta} with only one light neutrino $m_l$,  and one heavy neutrino $m_h$, respectively. The last two terms originate from interference between left- and right-handed interactions and have a distinct kinetic behavior with the additional factors $m_\upnu / E_\upnu = m_\upnu / ( E_0 - V_j - E )$ and $m_\up e / E_\up e = m_\up e / ( E + m_\up e )$. The interference terms for the light neutrino and the heavy neutrino have different signs, respectively, arising from the columns in the 2x2 mixing matrix. The coefficients are defined as

\begin{align}
    a_\up{LL} & = 1 + 2C \tan \xi \cos \alpha \ , \\
    a_\up{RR} & = \frac{m^4_{W_\up L}}{m^4_{W_\up R}} + \tan^2\xi 
    + 2C  \frac{m^2_{W_\up L}}{m^2_{W_\up R}} \tan \xi \cos \alpha \ , \\
    a_\up{LR} & = -2 \left(\frac{m^2_{W_\up L}}{m^2_{W_\up R}} + C \tan \xi \cos \alpha \right)
\end{align}

and

\begin{equation}
    C = \frac{g^2_V - 3 g^2_A}{g^2_V + 3 g_A^2} \simeq - 0.65 \ , 
\end{equation}

where $\xi$ is the mixing angle between left- and right-handed $W$ bosons and $\alpha$ a $CP$-violating phase. We furthermore neglect the possibility of complex phases involved in active-sterile mixing, so the coefficients for the LR term both for the sterile and the active neutrino differ only by the mass. Regarding the current experimental limits on these coefficients, the key observable is the right handed $W_\up R$ mass, which is connected with the mixing angle $\xi$ via the boson mixing matrix \cite{Barry2013}. The most robust bound comes from the LHC, which roughly states $m_{W_\up R} \gtrsim \SI{3}{TeV} $ \cite{ATLAS2012,CMS2014}. This translates into a bound on the LR mixing angle of about $|\xi| \lesssim \num{e-3}$. A theoretical lower bound on the CP violating phase has been derived in ref. \cite{Zhang2008}, stating $|\alpha| > 0.035$.

\subsection{Model-independent parametrization}

The theoretical spectrum \eqref{eq:beta-rh-theo} has two disadvantages for practical right-handed current searches in tritium $\upbeta$ decay. On one hand it is highly dependent on the underlying left-right symmetrical model. In this scenario a right-handed current contribution can only be present if there is active-sterile mixing, as can be seen in the $a_{LR}$ terms in eq. \eqref{eq:beta-rh-theo}. However, on different underlying theoretical considerations there can also be an identical signature without sterile contribution, e.g., by non-trivial scalar, pseudoscalar and tensor couplings \cite{Bonn2011, Stephenson1998}. On the other hand, the number of parameters in eq. \eqref{eq:beta-rh-theo} is higher than the degrees of freedom, which is problematic for a fit.

Thus, we are looking to transform eq. \eqref{eq:beta-rh-theo} in order to come up with a model-motivated, yet model-independent parametrization. We can then highlight the effective resulting shape of the $\upbeta$-spectrum. We do so by introducing an effective mixing angle $\theta\subeff$ through

\begin{align}
\label{eq:sin2eff}
    (a_\up {LL} + a_\up {RR}) \sin^2 \theta_\mathrm{eff} & = a_\up {LL} \sin^2 \theta +  a_\up {RR} \cos^2 \theta \\
\label{eq:cos2eff}
    (a_\up {LL} + a_\up {RR}) \cos^2 \theta_\mathrm{eff} & = a_\up {LL} \cos^2 \theta +  a_\up {RR} \sin^2 \theta \ .
\end{align}
Furthermore, the interference terms in eq. \eqref{eq:beta-rh-theo} can be parametrized by
\begin{equation}
\label{eq:crh}
c_\mathrm{LR} =  \frac{a_\up {LR}}{a_\up {LL} + a_\up {RR}}  \cdot \frac{m_\up e}{m_\up e + E_0}\ \cos \theta \sin \theta  \ ,
\end{equation}
where $m_\up e/(m_\up e + E)$ has been approximated by $m_\up e/(m_\up e + E_0)$ for a measurement near the endpoint. In the following, $c\subrm{LR}$ denotes the effective left-right interference strength. It can also be negative and acts effectively as a Fierz parameter. Note that while the interference strength is independent of the effective mixing angle as a fit parameter, it is still dependent on the \emph{physical} mixing angle. Since the values for $a_\up{LL}$ will be close to 1 and for $a_\up{LR}$ and $a_\up{RR}$  close to 0, the effective mixing angle will correspond roughly to the physical mixing angle. The resulting shape of the $\upbeta$-spectrum is then

\begin{equation}
\begin{split}
w(E) & = \sum \limits_j P_j \Bigg[ w_{h_j}'(E) \cdot  \sin^2 \theta_\mathrm{eff} 
 + w_{l_j}'(E) \cdot \cos^2\theta_\mathrm{eff} \\
 & + c_\mathrm{LR} \cdot \left( w_{h_j}'(E)  \cdot  \frac{m_h}{E_0-E}
- w_{l_j}'(E)  \cdot \frac{m_l}{E_0-E} \right) \Bigg] \ ,
\end{split}
\label{eq:beta-rh-eff}
\end{equation}

where the global factor $(a_{LL} + a_{RR})$ has been absorbed into the decay amplitude:

\begin{align}
	w_{l_j}'(E) & =  (a_\up{LL} + a_\up{RR}) \cdot w_{l_j}(E) \\
	w_{h_j}'(E) & =  (a_\up{LL} + a_\up{RR}) \cdot w_{h_j}(E) \ .
\end{align}

As outlined in the last section it may be possible that an effect on the $\upbeta$-spectrum with the same shape as the mixed terms in eq. \eqref{eq:beta-rh-eff} might be produced by a mechanism not based on left-right symmetry which can then be independent of the sterile mixing angle. The re-parametrized spectrum is model-agnostic and fits to a complete class of theoretical scenarios which predict the same term $\propto m_\upnu / E_\upnu$ with the effective Fierz parameter $c\subrm{LR}$  \cite{Stephenson2000}.

\subsection{Discussion of shape and parameter dependencies}

\begin{figure}[bt]
\centering
  \includegraphics[width=0.8\textwidth]{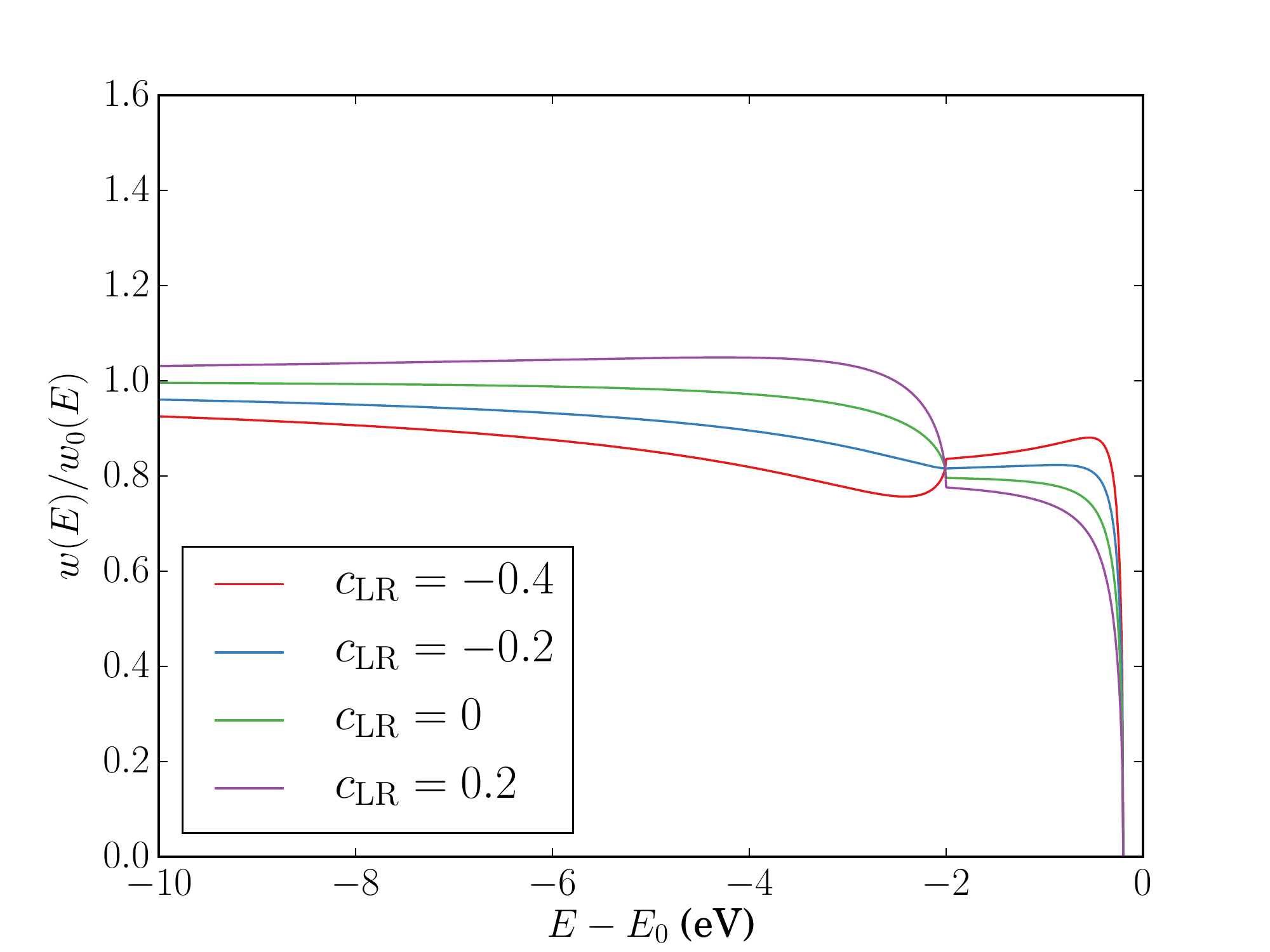}
  \caption{$\upbeta$-spectrum ratio $w(E)/w_0(E)$ near the endpoint  for different left-right interference strengths $c_\mathrm{LR}$ for effective mixing angle $\sin^2\theta_\mathrm{eff} = 0.2$, sterile neutrino mass $m_h = 2$ eV and light neutrino mass $m_l = 0.2$ eV. For simplicity, only the $V_j = 0$ component of the $\upbeta$-spectrum has been used. }
\label{fig:eff-strengths}
\end{figure}

\begin{figure}[bt]
\centering
  \includegraphics[width=0.8\textwidth]{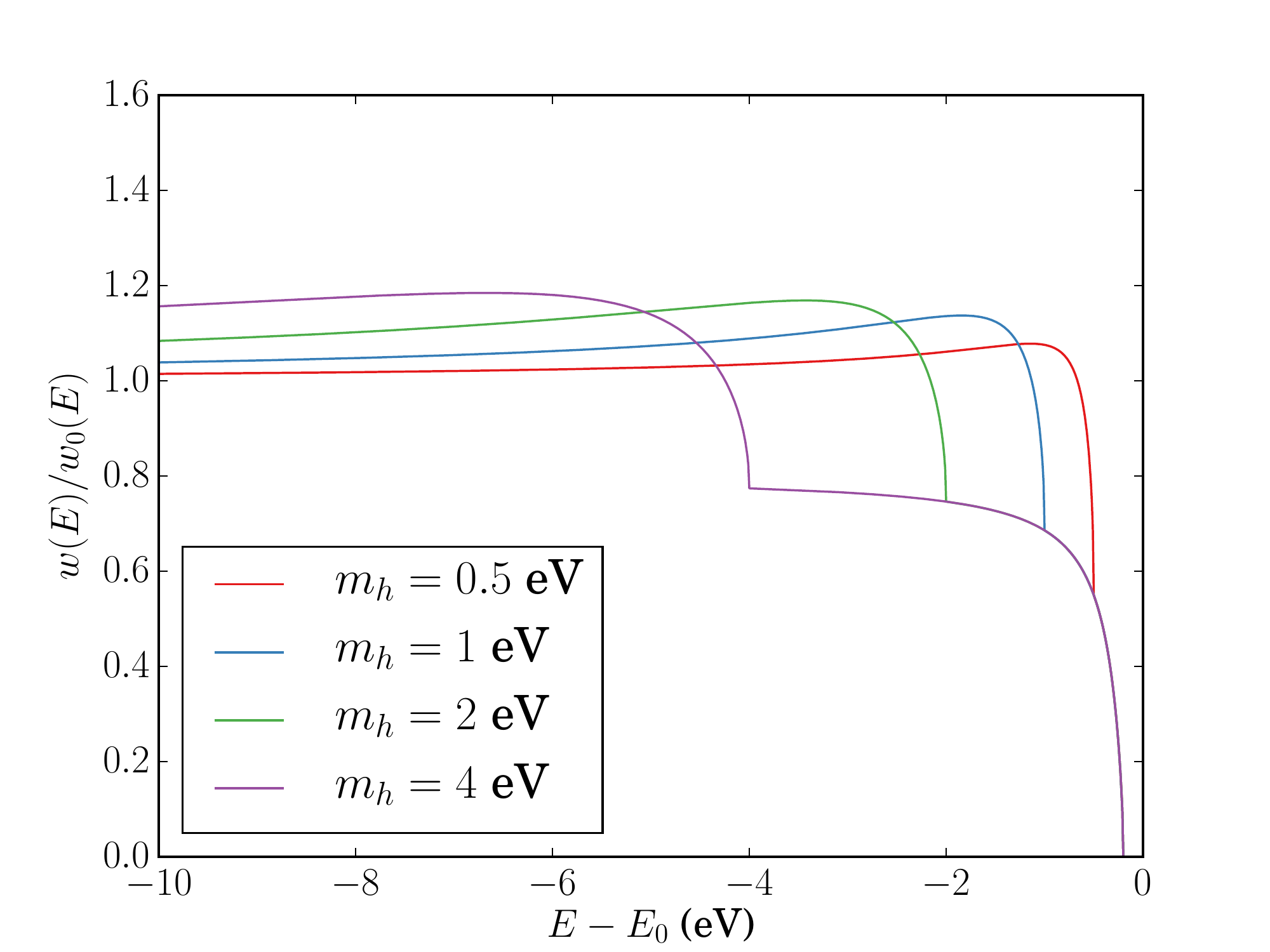}
  \caption{$\upbeta$-spectrum ratio $w(E)/w_0(E)$ near the endpoint for effective mixing angle $\sin^2\theta_\mathrm{eff} = 0.2$, left-right interference strength $c_\mathrm{LR} = 0.5$ and light neutrino mass $m_l = 0.2$ eV. For simplicity, only the $V_j = 0$ component of the $\upbeta$-spectrum has been used.}
\label{fig:sterile-masses}
\end{figure}

Since we are interested in the effective shape of the spectrum with right-handed currents \eqref{eq:beta-rh-eff} in relation to the standard $\upbeta$-spectrum \eqref{eq:beta}, we study the expression $w(E) / w_0(E) $, where $w_0(E)$ is the $\upbeta$-spectrum \eqref{eq:beta-rh-eff} with zero neutrino masses $m_l = m_h = 0$. For simplicity, only the electronic ground state $V_i = 0$ in \eqref{eq:beta} has been taken into consideration. It is plotted as a function of the energy for different effective left-right interference strengths $c\subrm{LR}$ in fig. \ref{fig:eff-strengths}. In case of $c\subrm{LR} = 0$, no contribution from left-right interference is present and the spectrum is purely a superposition of two $\upbeta$-spectra with neutrino masses, $m_l$ and $m_h$, respectively, according to eq. \eqref{eq:beta-rh-eff}. The exemplary contribution of the sterile mass state $m_h=\SI{2}{eV}$ with a strength of $\sin^2\theta=0.2$ in the plot leads to the well-known step-like feature at $E_0-m_h$ \cite{Formaggio2011, Riis2011}. A non-vanishing $c\subrm{LR}$ leads either to a boost or a suppression in the regions just below $E_0 - m_l$ and $E_0 - m_h$, respectively, spanning a few eV at maximum and depending on the sign of $c\subrm{LR}$. Due to the different sign of active and sterile interference terms \eqref{eq:beta-rh-eff}, a positive $c_{LR}$ causes a boost below $E_0 - m_h$, due to the term proportional to $m_h/(E_0 - E)$, together with a suppression below $E_0 - m_l$, due to the term proportional to $m_l/(E_0 - E)$, and vice versa for negative $c_{LR}$. The effect is slightly more pronounced below $E_0 - m_h$ than below $E_0 - m_l$ because of the proportionality to $m_h$ and $m_l$, respectively. To demonstrate the mass-dependency, the same expression is plotted in fig.~\ref{fig:sterile-masses} for a fixed $c\subrm{LR} = 0.5$ as a function of the sterile neutrino mass $m_h$. It can be seen that the peak magnitude of the boost increases only subtly with growing $m_h$, but the boost region, i.e. where $w(E)/w_0(E) > 1$, spans over an increasingly larger energy interval due to the proportionality to $m_h/(E_0 - E)$. In reality, the signature is washed out to some extent by the final state distribution in eq. \eqref{eq:beta}. Furthermore, in a tritium $\upbeta$-decay experiment using an integrating spectrometer, such as KATRIN, the differential $\upbeta$-decay spectrum is not accessible directly. Instead, the integral $\upbeta$-decay spectrum is measured, where the differential spectrum is convolved with the transmission function of the KATRIN main spectrometer \cite{Picard1992, KATRINDesignReport}. This is combined with further experimental corrections, such as inelastic scattering of electrons in the source, in order to define a response function, which will be looked at in more detail in the next section.

\section{KATRIN sensitivity}

In the following section we will apply the derived knowledge on the signature of right-handed currents with sterile neutrinos to the experimental parameters of the KATRIN experiment \cite{KATRINDesignReport} in order to estimate its sensitivity.

\subsection{Prerequesites}

\subsubsection*{KATRIN set-up and response}

KATRIN is designed to measure the electron neutrino mass $m_l$ with a sensitivity of \SI{0.2} {eV} at 90\% confidence level (CL) \cite{KATRINDesignReport}. The tritium is provided by a windowless gaseous molecular tritium source (WGTS) \cite{Babutzka2012} with a high activity of $\sim \SI{e11}{Bq}$. The electrons from the $\upbeta$ decay are filtered in the main spectrometer based on the \emph{magnetic adiabatic collimation with electrostatic filter (MAC-E Filter)} principle \cite{Picard1992}. The electrons are moving adiabatically from a high magnetic field at the tritium source with $B_\mathrm{source} = \SI{3.6}{T}$  into a small field of $B_\up A = \SI{0.3}{\milli\tesla}$ in the center. Since the relativistic magnetic moment is constant under adiabatic conditions, the electron momenta become aligned with the field. By additionally applying an electrostatic retarding potential $U$ in the center of the spectrometer, the main spectrometer acts as a high-pass filter with a sharp energy resolution of $\Delta E / E = B_\up A / B_\mathrm{max}$, that is, $\Delta E \approx \SI{0.93}{eV}$  at $E_0 = \SI{18.6}{keV}$\footnote{An additional pinch magnet with $B$ field $B\submax=\SI{6}{T}$ restricts the maximum starting angle w.r.t. the source to $\theta\submax = \arcsin \sqrt{\frac{B\subrm{source}}{B\submax}}$.}.  In the focal plane detector (FPD) the count rate is then measured as a function of $qU$, thus effectively integrating the $\upbeta$-spectrum between $qU$ and $E_0$, where $q$ denotes the negative electron charge.

\begin{figure}[h!bt]
\centering
  \includegraphics[width=0.8\textwidth]{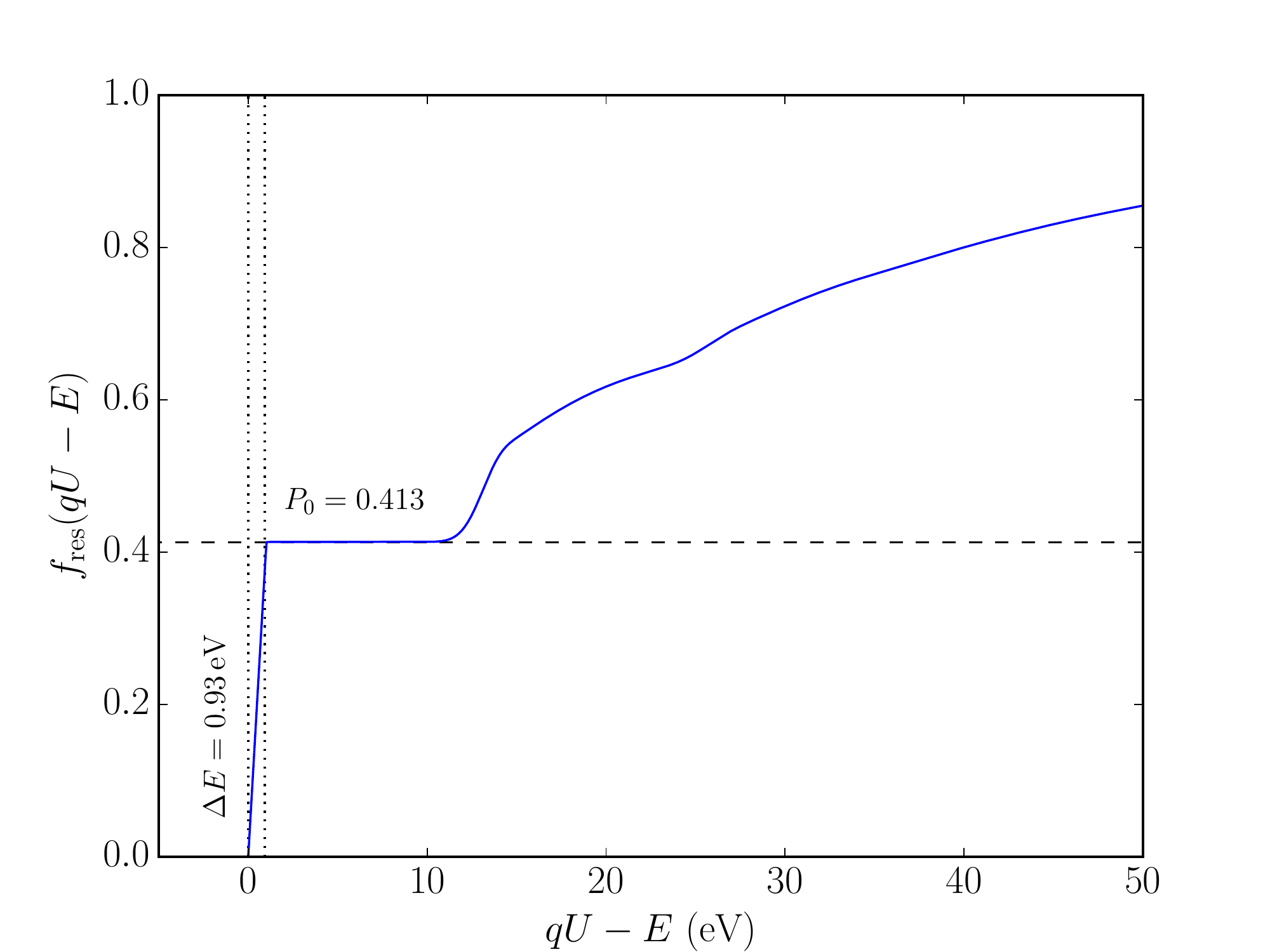}
  \caption{Response function of the KATRIN experiment for isotropically emitted electrons \eqref{eq:response} close to the endpoint $E_0$. The first $\sim$ 10 eV are described by the transmission function $T(E-qU)$ of a MAC-E-Filter \cite{Picard1992} of width $\SI{0.93}{eV}$, leading into a plateau given by the fraction of electrons of $ \sim 41\%$ that have undergone no inelastic scattering process. The region with $qU - E \gtrsim \SI{10}{eV}$ is increasingly dominated by electrons with larger start energies who lose a certain amount of energy in inelastic collisions in the tritium source.}
  \label{fig:response}
\end{figure}

However, in addition to the energy resolution of the main spectrometer, experimental effects directly affecting the energy distribution of the electrons also need to be considered, such as inelastic scattering in the source. The integral count rate is therefore given by

\begin{equation}
  \label{eq:betaint}
  R(qU) = \epsilon \cdot \frac{\Delta\Omega}{4\pi} ~ \left(\int_{qU}^{E_0} \diff E ~ w(E) \cdot f\subrm{res}(E - qU) ~ \right) + b ,
\end{equation}
where $\frac{\Delta\Omega}{4\pi} =  (1 - \cos \theta\subrm{ max} )/2$ is the accepted solid angle with $\theta\submax = \SI{50.77}{\degree}$ and $b$ the background rate. A factor $\epsilon$ is taken into account to model various approximately energy-independent losses, which are in this case the fraction of the transmitted flux tube of the WGTS, $\epsilon\subrm{flux} \approx 0.83$, and the detector efficiency $\epsilon\subrm{det} \approx 0.9$. The response function is given by

\begin{equation}
  \label{eq:response}
  f\subrm{res}(E - qU) = T(E - qU) \otimes f\subrm{loss}(E),
\end{equation}
where $T(E - qU)$ is the analytical transmission function for isotropic electrons \cite{KATRINDesignReport} of width $\SI{0.93}{eV}$, defined by the field configuration, and $f\subrm{loss}(E)$ the energy loss spectrum by inelastic scattering in the source \cite{Aseev2000, Hannen2017} for a column density of $\rho d = \SI{5e17}{\per\cm^2}$ and averaged over the starting angle $\theta$ for isotropic emission. The  response function \eqref{eq:response} is plotted in fig. \ref{fig:response}.

\begin{figure}[hbt]
\centering
  \includegraphics[width=0.69\textwidth]{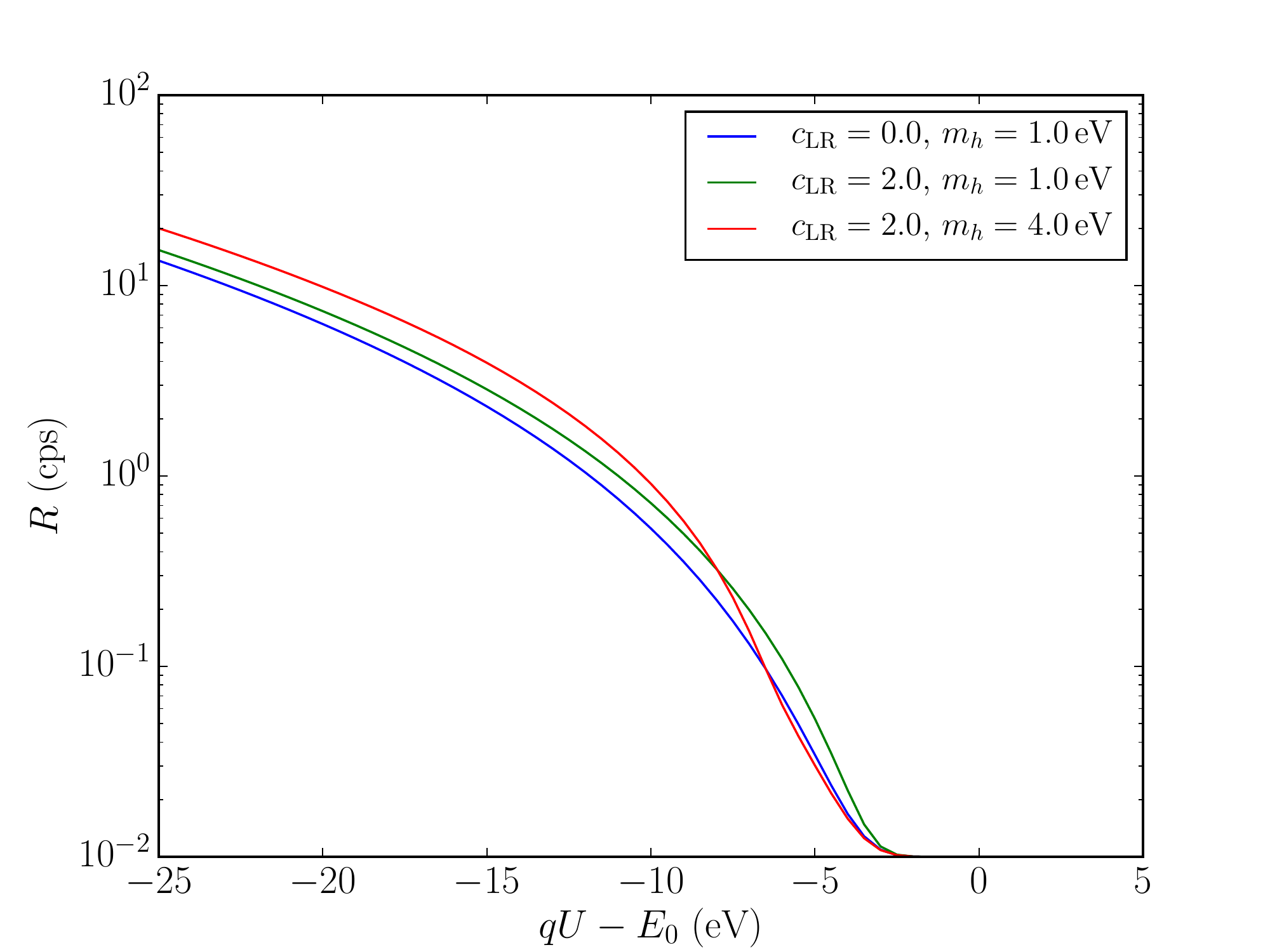}
  \includegraphics[width=0.69\textwidth]{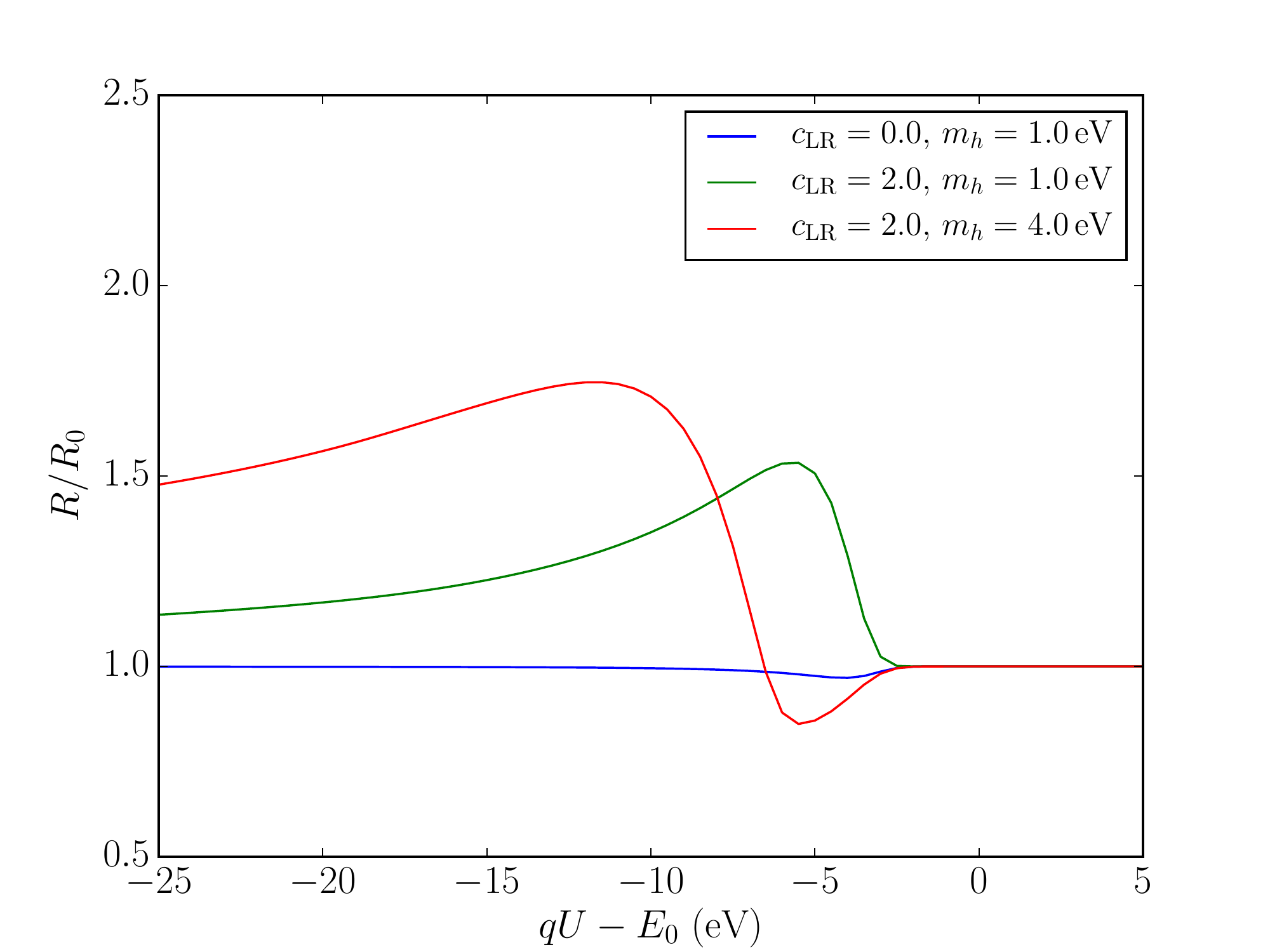}
  \caption{Integral spectrum $R(qU-E_0)$ of KATRIN \eqref{eq:betaint} (upper plot) for different exemplary scenarii with exaggerated effective~left-right interference strength $c\subrm{LR}$ and different sterile neutrino masses $m_h$. The effective mixing is $\sin^2\theta = 0.2$ and the electron neutrino mass $m_l = 0$, plus a uniform background of $b = \SI{10}{mcps}$. The lower plot shows the ratio $R/R_0$, where $R_0$ is the integral $\upbeta$-spectrum with $c\subrm{LR}=0$ and $m_h=0$. The endpoint is smeared and effectively lowered due to rotational-vibrational excitations of the daughter molecule with an average energy of $\SI{1.7}{eV}$ \cite{Saenz2000}. The effect of a positive $c\subrm{LR}$ can be seen as a boost in a region below the endpoint, where the strength and average stretch of the boost is determined by the sterile mass $m_h$. The relative suppression at $\sim \SI{-5}{eV}$ with the red curve is an effect of $\cos^2\theta\subeff<1$, visible for sufficiently large $m_h$.}
  \label{fig:integral}
\end{figure}

The resulting integral spectrum \eqref{eq:betaint} is shown in fig.~\ref{fig:integral} for different sterile neutrino masses and effective left-right interference strengths. The final state distribution has been taken from \cite{Saenz2000} and the energy loss spectrum from \cite{Aseev2000}. Due to integration in combination with these experimental effects, the signature is clearly weaker than in the differential spectrum \eqref{eq:beta-rh-eff}. Nevertheless, for large enough values of $c_\up{LR}$ there is still a distinct effect on the shape, sufficiently large to be detected with the high precision experiment KATRIN. As within the differential $\upbeta$-spectrum, the existence of right-handed currents is manifest as a boost or suppression in a region close to the endpoint, for a positive or negative $c\subrm{LR}$, respectively. The strength of the signature increases with the sterile mass $m_h$ as well. Furthermore, the plot suggests that for a lower sterile neutrino mass the boost (or suppression) is more localized than for a higher sterile neutrino mass. In the curve corresponding to $m_h = \SI{4}{eV}$ shown in fig.~\ref{fig:integral}, the boost region clearly stretches down below the KATRIN default lower measurement interval bound of $qU = E_0 - \SI{25}{eV}$ \cite{KATRINDesignReport}.

\subsubsection*{Analysis method}

From the integral spectrum \eqref{eq:betaint} and consequently the likelihood shape, we want to derive the sensitivity on the left-right interference $c\subrm{LR}$ after the default measurement time period of three effective years (corresponding to five calendar years) with KATRIN.
The sensitivity of a parameter, in this context, is identified with the uncertainty or upper limit of its fit estimate with respect to a fiducial input value, given by the null hypothesis (i.e. $c\subrm{LR}=0$).
Due to a larger number of fit parameters in the scenario of added light sterile neutrinos and right-handed currents and the possibility of a complex fit parameter distribution with non-Gaussian errors and non-linear correlations, a Bayesian approach has been chosen instead of the common frequentist paradigm. This has been performed using a Markov Chain Monte Carlo (MCMC) analysis with the likelihood function

\begin{equation}
  \label{eq:likelihood}
  \log L(\theta) = - \frac 1 2 \sum_{i=1}^m \frac{(n_i(\Theta) - n_i(\Theta_0))^2}{n_i(\Theta)} ~ , 
\end{equation}
with variable parameters $\Theta$ and null-hypothesis $\Theta_0$. Such a likelihood function utilizes the null-hypothesis instead of toy data, but effectively approximates the posterior distribution of possible data-sets. The expected counts in each bin $n_i$ are given by the integral $\upbeta$-spectrum \eqref{eq:betaint} as

\begin{equation}
  \label{eq:counts}
  n_i = t_i \cdot R(qU_i) ~ , 
\end{equation}
with a set of $m$ measurement points $qU_i$ with measurement time $t_i$, respectively. For the measurement time distribution $t_i$ the proposed distribution for in \cite{KATRINDesignReport} with a lower interval bound of $qU_1 = E_0 - \SI{25}{eV}$ has been used, assuming that the main measurement objective of KATRIN will be the measurement of the active neutrino mass $m_l$. With the fit function \eqref{eq:counts}, the parameters of the model are

\begin{itemize}
	\item the $\upbeta$-decay spectral endpoint $E_0$,
	\item the active neutrino mass $m_l$,
	\item the sterile neutrino mass $m_h$,
	\item the effective mixing angle $\sin^2\theta\subrm{eff}$,
	\item the effective left-right interference strength $c\subrm{LR}$
	\item the decay amplitude $S$ (i.e. a normalization factor, defined by the integral count-rate \eqref{eq:betaint} for $qU=0$) and
	\item the background rate $b$.
\end{itemize}

The Differential Evolution Markov Chain Monte Carlo (DEMC) algorithm \cite{TerBraak2006} has been used for the simulations. DEMC is an ensemble-based MCMC method, where instead of a single Markov chain an ensemble of $N$ chains is run.  The proposal for each step in chain $j$ at iteration $t$ is generated by adding the difference of two other randomly selected chains, multiplied with a scaling parameter $\gamma$:

\begin{equation}
  \theta_{j, t+1}' = \theta_{j, t} + \gamma (\theta_{k, t} - \theta_{l, t}) \qquad j \neq k \neq l \ .
\end{equation}

The proposal is then accepted or rejected with the classic Metropolis Hastings criterion \cite{Hastings1970}. This scheme solves two problems with the classic Metropolis algorithm, which is the choice of the scale and the orientation for the proposal distribution. Instead of tuning these by hand, these are here implicitly derived from the ensemble at each iteration. This is especially useful in cases where the posterior distribution shows a high correlation between parameters. 

Regarding the scaling parameter, the recommendation by \cite{TerBraak2006} has been used in our implementation, which states $\gamma=2.38/\sqrt{(2D)}$, where $D$ is the dimension of the parameter space. In most cases this choice should provide an optimal acceptance ratio. The remaining tuning parameter is then the number of chains $N$, which should at least be $N = 2 D$ and needs to be increased for more complex posterior distributions. No explicit Bayesian prior has been chosen, only the physical parameter boundaries have been enforced which are in particular $m_l, m_h > 0$ and $0 < \sin^2\theta\subrm{eff} < 1$. To check convergence, the Gelman Rubin $R$ diagnostic \cite{Gelman1992} has been used, which needs to be close to 1. A common convergence criterion is e.g. $R<1.01$, which has been fulfilled in all cases, except with a free neutrino mass parameter (see below). All results have been cross-checked and could be reproduced with an independent simulation based on a classic Metropolis Hastings \cite{Hastings1970} algorithm without adaption. However, using the latter requires careful manual fine-tuning of the proposal distributions.

\subsection{Results}

\subsubsection*{Credible intervals}

\begin{figure}[hbt]
\centering
  \includegraphics[width=0.9\textwidth]{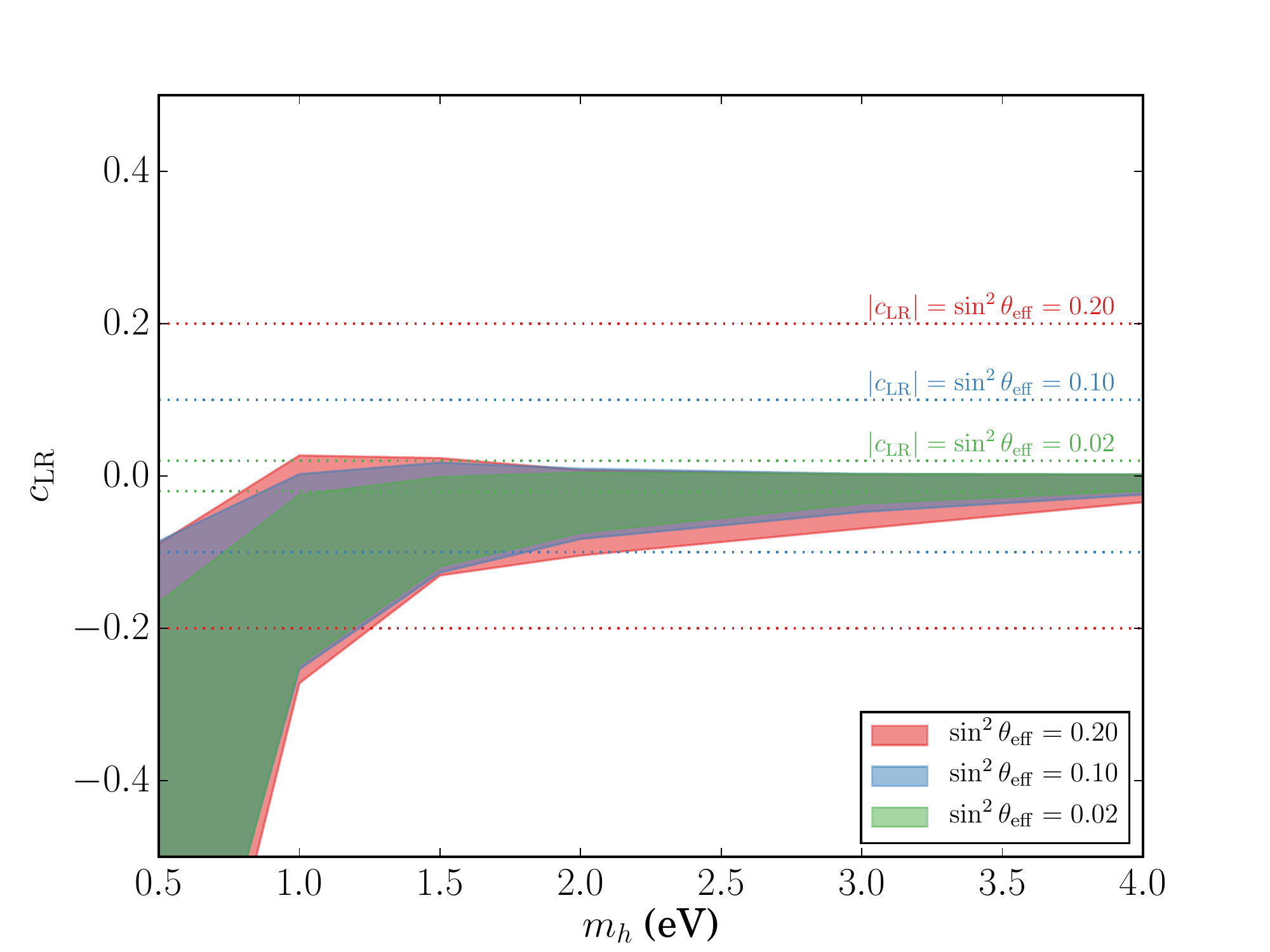}
  \caption{Bayesian credible intervals (95 \% credibility level) on null hypothesis \eqref{eq:likelihood} for effective left-right interference strength $c\subrm{LR}=0$ as a function of fixed sterile neutrino mass $m_h$ with different effective mixing angles $\sin^2 \theta\subrm{eff}$. The null hypothesis further includes $m_h=0$, $E_0 = \SI{18.575}{keV}$, $b=\SI{10}{mcps}$ and the KATRIN default signal amplitude. The dotted horizontal lines represent the hard limit of $|c\subrm{LR}| < \sin^2\theta\subrm{eff}$ in case of LR symmetry.}
  \label{fig:sensitivity-fixms}
\end{figure}

Figure~\ref{fig:sensitivity-fixms} shows the 95 \% credible interval of the effective left-right interference strength $c\subrm{LR}$, given a null-hypothesis of $c\subrm{LR} = 0$ in \eqref{eq:likelihood} for different sterile neutrino masses and effective mixing angles with light neutrino mass of $m_l=0$, background rate of $b=\SI{10}{mcps}$ and the KATRIN default signal amplitude. The sterile neutrino mass has been fixed in the MCMC runs, so the plot can be interpreted as statistical sensitivity on an excess at a certain mass $m_h$. The results are based on DEMC runs with 20000 iterations in each chain and an ensemble size $N=4D$. Several pieces of information can be extracted from the plot. The average width of the interval varies from about 0.5 to 0.05 in terms of $c\subrm{LR}$. One of the most distinct features is a strong bias. For a mass of $m_h=\SI{0.5}{eV}$ the null-hypothesis is not even in the 95 \% credible interval. 
There is no a priori reason to not expect a bias. It can be attributed to a \emph{volume effect} in the space of the posterior. While the point of maximum likelihood is indeed identical with the fiducial point $\Theta_0$ (within a small numerical uncertainty), the marginalized posterior in the $c\subrm{LR}$ subspace has its maximum at $c\subrm{LR} \neq 0$, since it is integrated over all other dimensions. This will be looked at in more detail in the next subsection.

Besides that, a very strong dependence on the mass of the sterile neutrino can be seen, where the sensitivity improves drastically for heavier sterile neutrinos. This is prima facie a consequence of the proportionality of the left-right mixing terms in eq. \eqref{eq:beta-rh-eff} to the neutrino mass. Apart from that, for small sterile neutrino masses the interference terms for the active and sterile part cancel each other partly since they have opposite signs. Furthermore, the information about left-right interference, active-sterile mixing and the active neutrino mass is distributed in a broader region of the spectrum if the sterile neutrino mass is heavier. The sensitivity is less dependent on the effective mixing angle. However, smaller mixing angles seem to be slightly favorable. This is plausible, since the left-right interference term in eq. \eqref{eq:beta-rh-eff} depends only on the sterile mass, not on the effective mixing angle. A smaller effective mixing angle is thus expected to lead to a slightly clearer right-handed current signature. Nevertheless, for each effective mixing angle there is also a theoretical boundary for $c\subrm{LR}$, if LR symmetry is assumed. From eqs.~\eqref{eq:sin2eff}, \eqref{eq:cos2eff} and \eqref{eq:crh} we can conclude a hard limit $|c\subrm{LR}| < \sin^2\theta\subrm{eff}$, which is also shown in fig. \ref{fig:sensitivity-fixms}. 

\begin{figure}[h!bt]
\centering
  \includegraphics[width=\textwidth]{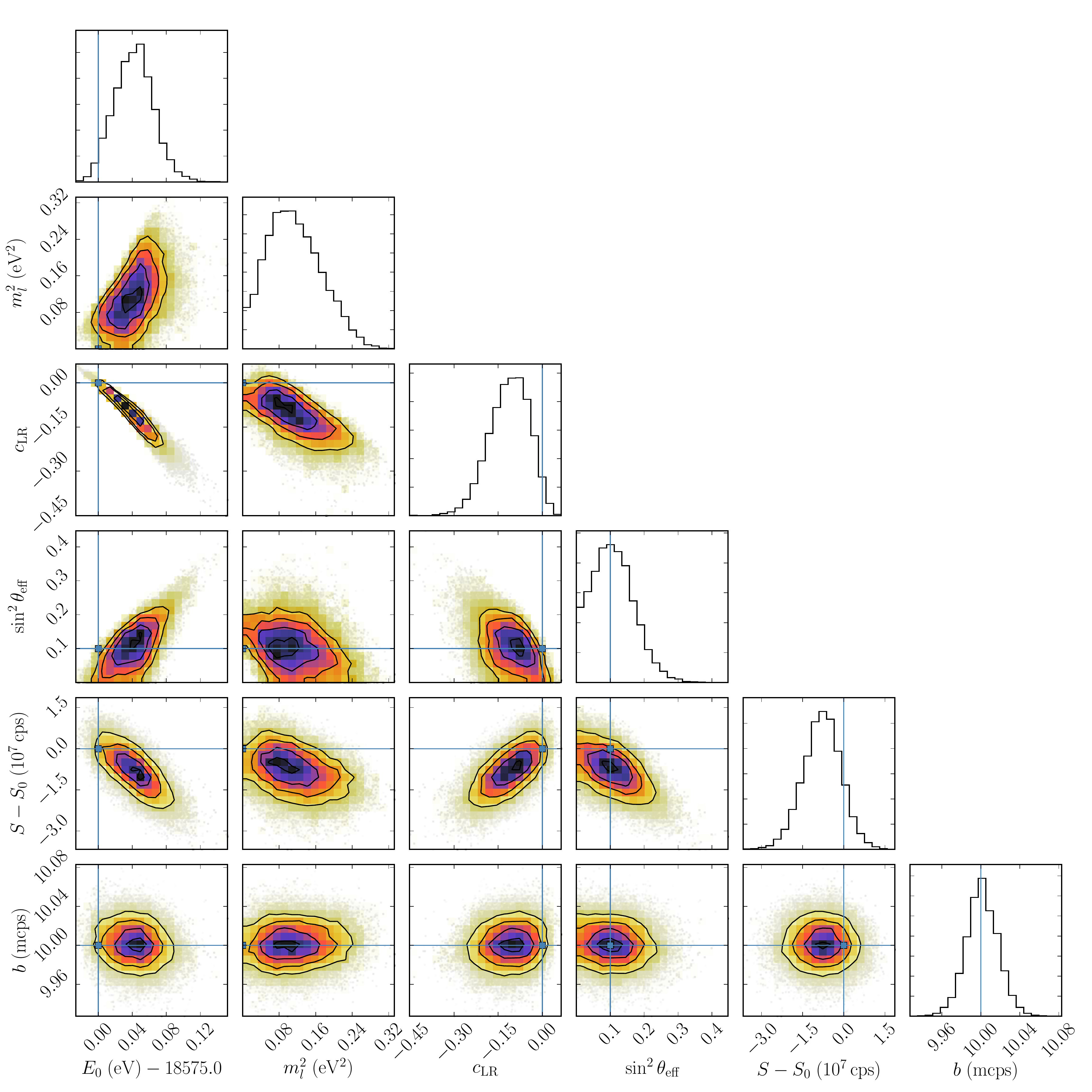}
  \caption{Marginalized posterior distributions for MCMC run with fixed sterile neutrino mass $m_h = \SI{1}{eV}$, effective mixing angle $\sin^2 \theta\subrm{eff} = 0.1$ and left-right interference strength $c\subrm{LR}=0$ for all combinations of the free fit parameters used. Contours are 0.5, 1, 1.5 and 2 $\sigma$, respectively. The blue lines show the fiducial values. Strong correlation between effective left right interference strength $c\subrm{LR}$ and $\upbeta$-decay endpoint $E_0$ can be observed. The Gelman Rubin statistic $R$ is well below 1.01 for all parameters.}
  \label{fig:fixms-corner1}
\end{figure}

\begin{figure}[h!bt]
\centering
  \includegraphics[width=\textwidth]{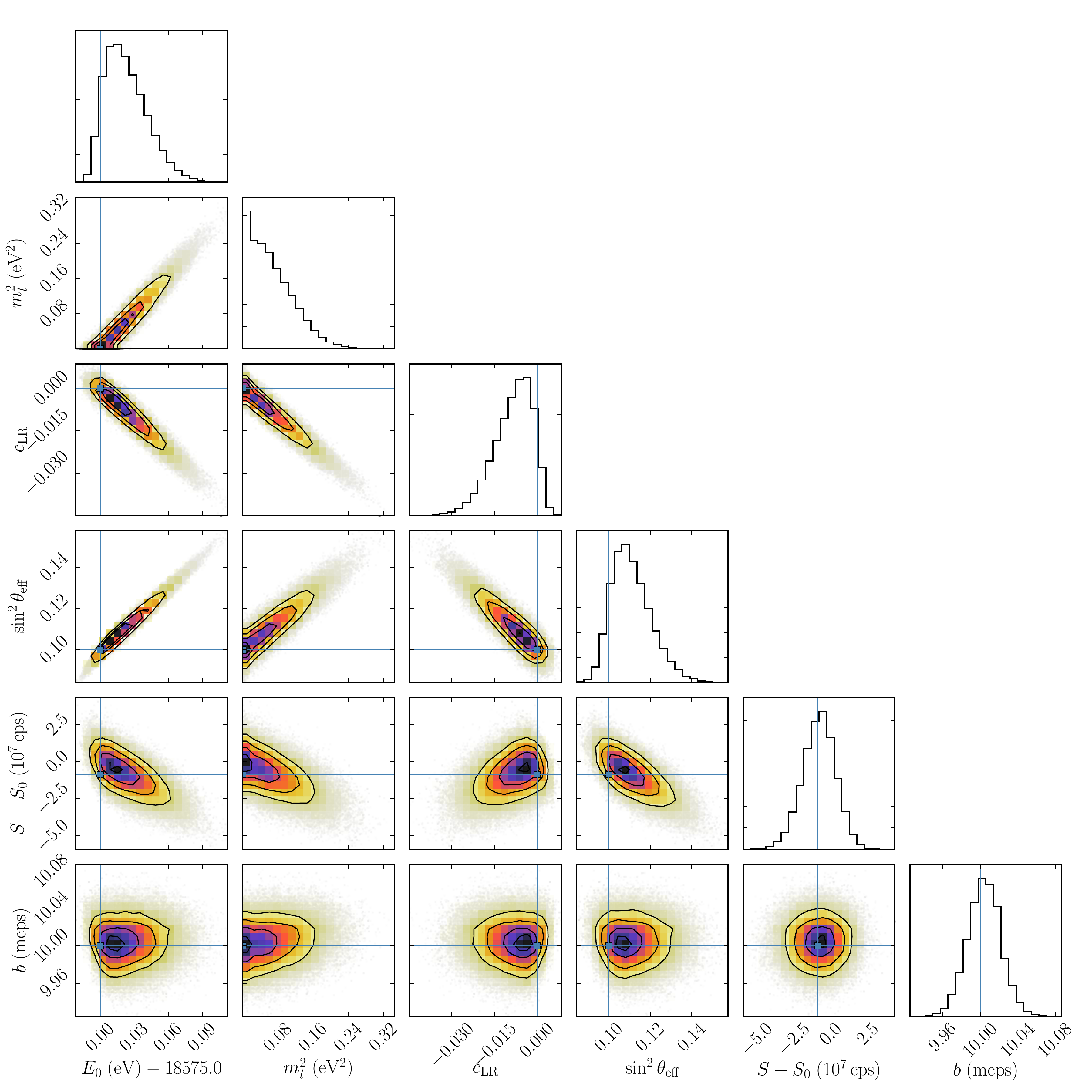}
  \caption{Marginalized posterior distributions for MCMC run with fixed sterile neutrino mass $m_h = \SI{4}{eV}$, effective mixing angle $\sin^2 \theta\subrm{eff} = 0.1$ and left-right interference strength $c\subrm{LR}=0$ for all combinations of the free fit parameters used. Contours are 0.5, 1, 1.5 and 2 $\sigma$, respectively.  The blue lines show the fiducial values. The correlation between effective left right interference strength $c\subrm{LR}$ and $\upbeta$-decay endpoint $E_0$ becomes slightly weaker. The correlations between $c\subrm{LR}$ and the sterile neutrino parameters $\sin^2 \theta\subrm{eff}$ and $m_h$ as well as the light neutrino mass $m_l$ become more distinct. However, with all correlated parameters the uncertainty decreases. The Gelman Rubin statistic $R$ is well below 1.01 for all parameters.}
  \label{fig:fixms-corner4}
\end{figure}

\subsubsection*{Parameter correlations}

A closer understanding of these observations can be accomplished by studying the parameter correlations in the posterior distribution. Figures \ref{fig:fixms-corner1} and \ref{fig:fixms-corner4} show the marginalized posterior distributions for $\sin^2 \theta\subrm{eff} = 0.1$ with $m_h = \SI{1}{eV}$ and $m_h = \SI{4}{eV}$, respectively. 

Two things are worth noticing. First, the distributions are significantly broader in the $m_h = \SI{1}{eV}$ case for all parameters except the background rate. Especially the effective mixing angle reaches zero in a large number of samples. This suggests that for a low sterile neutrino mass, the signatures of an active neutrino, a sterile neutrino and the right-handed current, along with an unknown endpoint, are too close to be distinguished. This is different in the case of $m_h = \SI{4}{eV}$, where more distinct linear correlations can be seen, but in total the parameter signatures are well distinguishable. This argument is sound in light of the findings from the last section, where a larger sterile neutrino mass has been shown to increase both the strength and the width of the right-handed current signature.  

Second, there is a strong linear correlation between the endpoint and the left-right interference strength. This correlation is weakened in the case of $m_h = \SI{4}{eV}$ and the point of highest density is closer to the fiducial null-hypothesis value (blue lines). Especially in the latter case there is a significant asymmetry which favors a lower left-right interference strength $c\subrm{LR}$ and a higher endpoint. That means most likely that changes in this direction are possible to compensate by choices of the other parameters but not vice versa. This leads to the supposition that endpoint-interference-correlation plays a central role in the volume effect, leading to the bias observed in fig.~\ref{fig:sensitivity-fixms}.

\subsubsection*{Fixed endpoint}

\begin{figure}[h!bt]
\centering
  \includegraphics[width=0.9\textwidth]{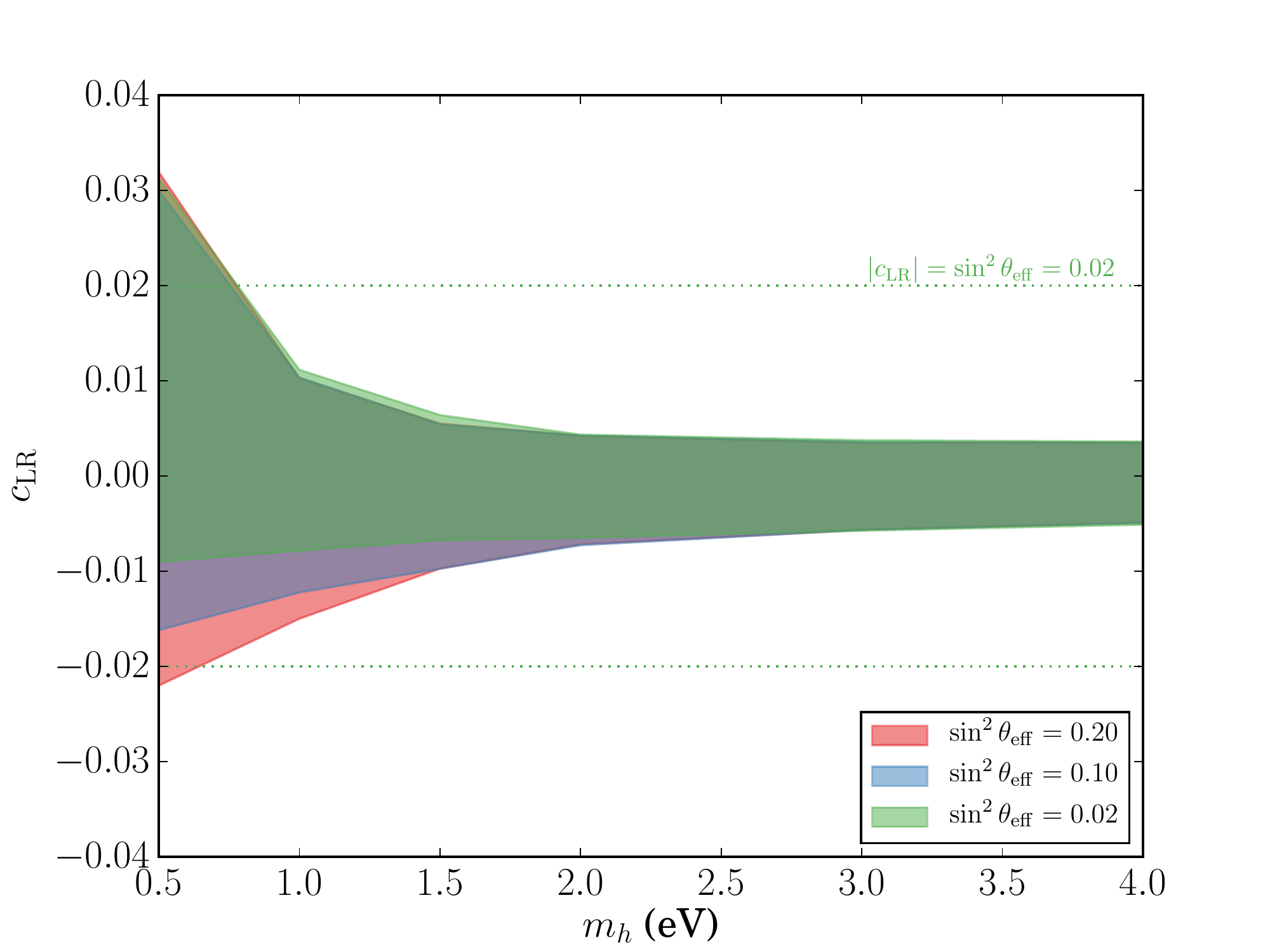}
  \caption{Bayesian credible intervals (95 \% credibility level) on null hypothesis \eqref{eq:likelihood} for effective left-right interference strength $c\subrm{LR}=0$ as a function of fixed sterile neutrino mass $m_h$ with different effective mixing angles $\sin^2 \theta\subrm{eff}$. The endpoint has been fixed at $E_0 = \SI{18.575}{keV}$. The fixation of the endpoint causes the bias largely to disappear.}
  \label{fig:sensitivity-fixe0}
\end{figure}

\begin{figure}[h!bt]
\centering
  \includegraphics[width=0.9\textwidth]{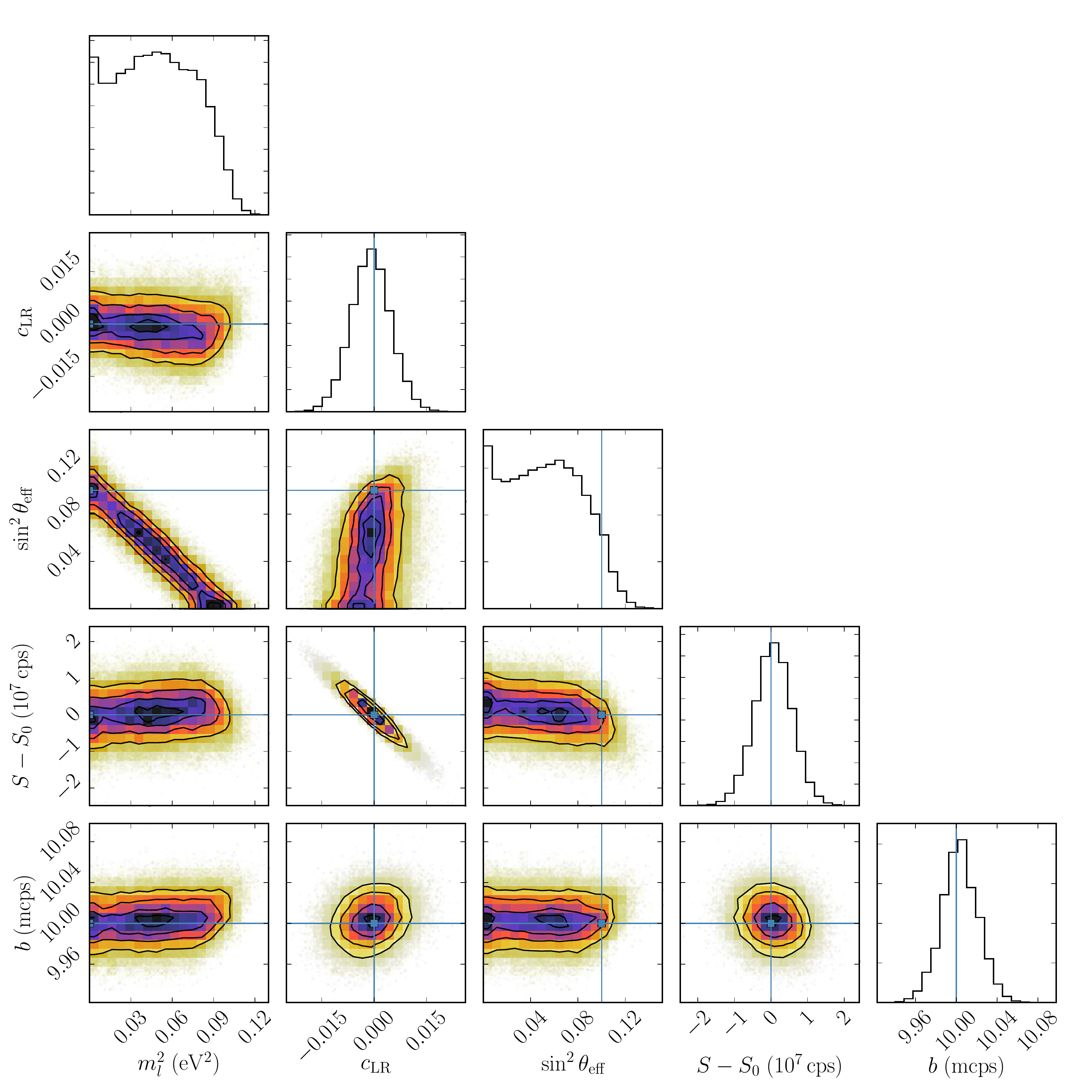}
  \caption{Marginalized posterior distributions for MCMC run with fixed sterile neutrino mass $m_h = \SI{1}{eV}$, effective mixing angle $\sin^2 \theta\subrm{eff} = 0.1$, left-right interference strength $c\subrm{LR}=0$ and a fixed endpoint $E_0 = \SI{18.575}{keV}$ for all combinations of the free fit parameters used. Contours are 0.5, 1, 1.5 and 2 $\sigma$, respectively. The blue lines show the fiducial values. The Gelman Rubin statistic $R$ is well below 1.01 for all parameters.}
  \label{fig:fixe0-corner1}
\end{figure}

The influence of the endpoint-interference correlation on the bias has been confirmed by repeating the simulation with a fixed endpoint, as shown in fig.~\ref{fig:sensitivity-fixe0}. The bias is reduced to a minimum. Further, the credible intervals narrow significantly, nearly by an order of magnitude. The sensitivity is still slightly better without sterile contribution, but not significantly. Fig. \ref{fig:fixe0-corner1} shows the corresponding marginal posterior distributions for $m_h = \SI{1}{eV}$. There is still a big uncertainty on the effective mixing angle. It continues to be correlated with the active neutrino mass, where the spectral shape consistent with the null hypothesis of an effective mixing $\sin^2\theta\subrm{eff} = 0.1$ can also be interpreted as a non-vanishing active neutrino mass. However, the distribution in the $c\subrm{LR}$-space is now unbiased and symmetric, since no other parameter choices are now possible any more which would fake a right handed current signature.

\subsubsection*{Constrained endpoint}

That leads to the question, if it is possible to constrain the endpoint by external measurements \cite{Streubel2014}, how such a constraint will quantitatively influence the sensitivity. To this end, the initial likelihood function \eqref{eq:likelihood} is modified by a prior on $E_0$:

\begin{equation}
  \label{eq:endpointconstraint}
  \log L'(\Theta) =  \log L(\Theta) - \frac 1 2\frac{(\left<E_0\right> - E_0)^2}{\Delta E_0^2} ~ , 
\end{equation}
where $\left<E_0\right>=\SI{18.575}{keV}$ is the null-hypothesis value and $\Delta E_0$ is the one $\sigma$ uncertainty on $E_0$.

\begin{figure}[h!bt]
\centering
  \includegraphics[width=0.9\textwidth]{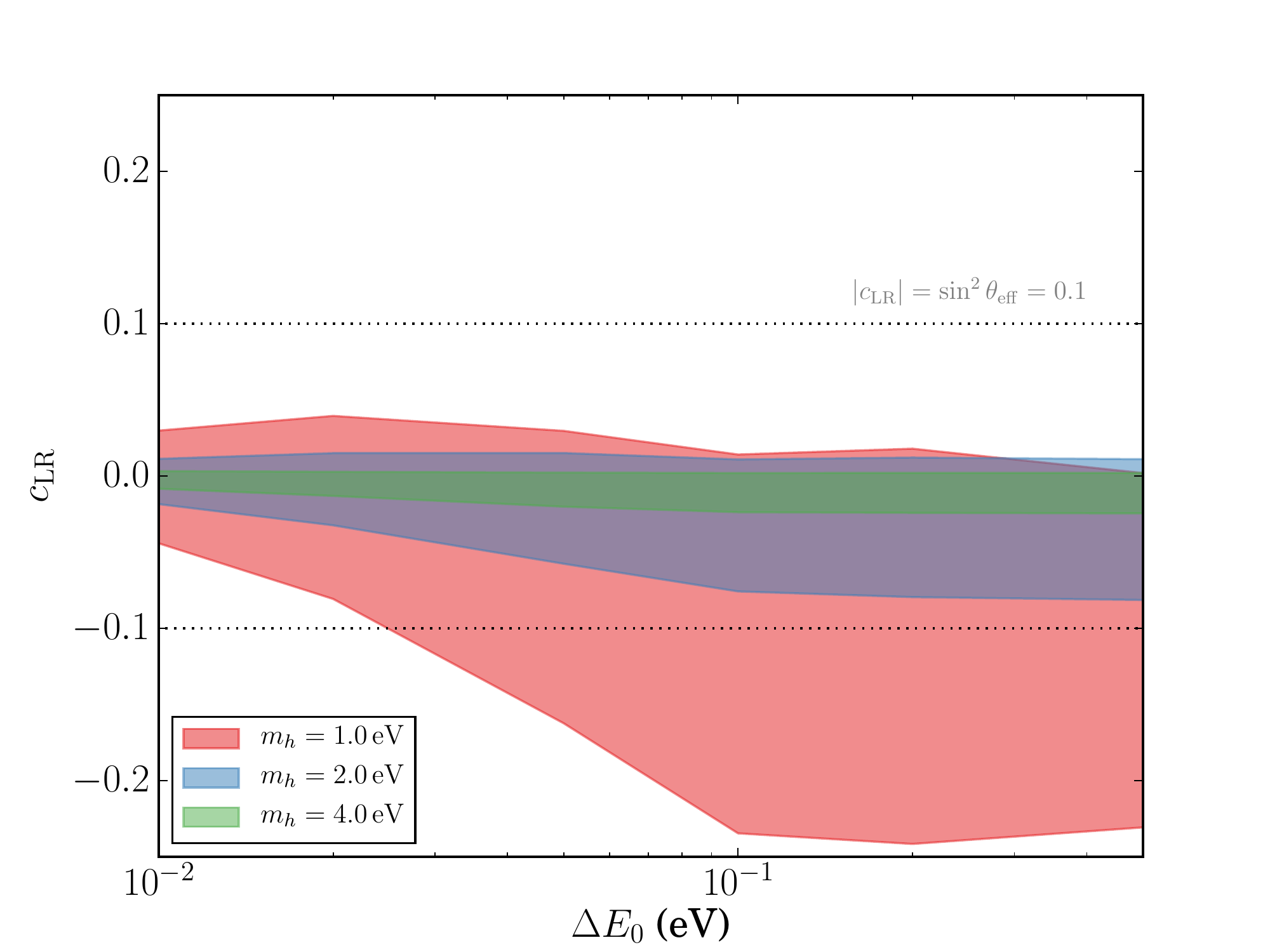}
  \caption{Bayesian credible intervals (95 \% credibility level) on null hypothesis \eqref{eq:likelihood} for effective left-right interference strength $c\subrm{LR}=0$ and effective mixing angle $\sin^2 \theta\subrm{eff} = 0.1$ with constrained endpoint \eqref{eq:endpointconstraint} as a function of one $\sigma$ endpoint uncertainty $\Delta E_0$ with fixed sterile neutrino masses $m_h$.}
  \label{fig:e0pull-sens}
\end{figure}

 In fig. ~\ref{fig:e0pull-sens}, the credible intervals are shown as function of the one $\sigma$ endpoint uncertainty $\Delta E_0$  for different sterile masses $m_h$. It can be seen that with decreasing endpoint uncertainty, the bias is reduced and the sensitivity increased. This is especially clear for lower sterile neutrino masses. There is however no significant effect unless the constraint exceeds \SI{0.1}{eV} precision. Current $Q$ value bounds from $^3$H-$^3$He mass measurements with precision Penning traps \cite{Myers2015} can be translated into a constraint of $\sim \SI{0.1}{eV}$. Future experiments aim for a bound of $\sim \SI{30}{meV}$ \cite{Streubel2014}. However, molecular effects and nuclear recoil have to be taken into account, which can possibly weaken the constraint on the endpoint \cite{Bodine2015}.

\subsubsection*{Sterile neutrino mass as free parameter}

\begin{figure}[h!bt]
\centering
  \includegraphics[width=0.95\textwidth]{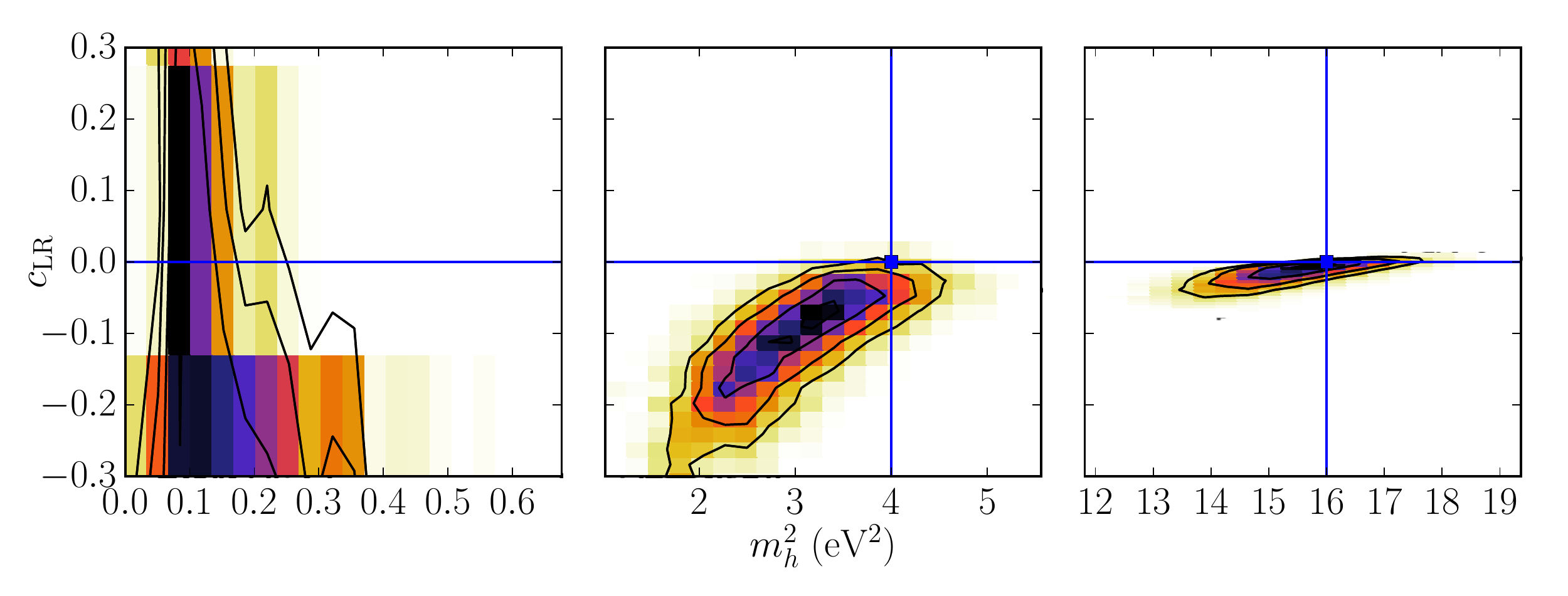}
  \includegraphics[width=0.95\textwidth]{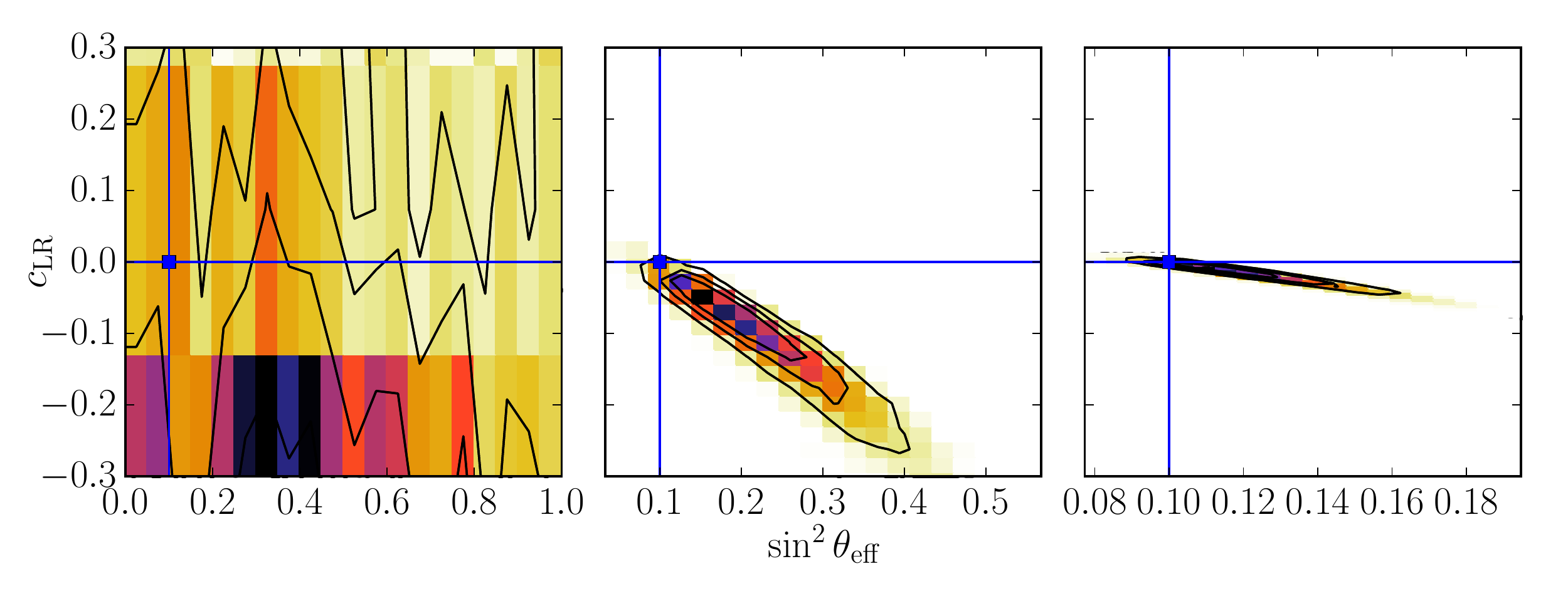}
  \caption{Selected marginalized posterior distributions for MCMC run with free sterile neutrino mass parameter and free endpoint, using fiducial values $m_h = \SI{1}{eV}$ (left), $\SI{2}{eV}$ (middle), $\SI{4}{eV}$ (right) and $\sin^2 \theta\subrm{eff} = 0.1$ (all). Contours are 0.5, 1, 1.5 and 2 $\sigma$, respectively. Upper panel: effective left-right interference strength vs.~squared sterile neutrino mass $m_h^2$; lower panel: effective left-right interference strength vs.~effective mixing angle. The blue lines show the fiducial values. Clearly, the fiducial mass and effective mixing angle fail to be estimated for smaller sterile neutrino masses $m_h \lesssim \SI{2}{eV}$. Only the chains in the case of $m_h^2 = \SI{4}{eV}$ are well converged with a Gelman Rubin statistic of $R < 1.01$.}
  \label{fig:panel-varms}
\end{figure}

In the simulations presented up to now, the sterile neutrino mass has been fixed. This has been motivated by the degeneracy one runs into when $c_\mr{LR}$ and $\sin^2\theta_\mr{eff}$ become small. This is a valid strategy for an exclusion, where the upper and lower limits, respectively, on these parameters can be determined as a function of the mass. However, in case a non-vanishing effective mixing angle is measured, the sterile neutrino mass either needs to be put in externally or treated as a free fit parameter in order to determine the correct credible intervals for $c\subrm{LR}$ and $\sin^2\theta\subrm{eff}$. While for a model without right-handed currents, KATRIN is able to test the sterile neutrino parameter space favored by the reactor antineutrino anomaly \cite{Riis2011}, the additional degeneracy brought in by the free parameter $c\subrm{LR}$, makes the situation more complicated. Fig. \ref{fig:panel-varms} shows the marginalized posterior distribution in the $(c\subrm{LR}, m_h^2)$-space (upper panel) and the $(c\subrm{LR}, \sin^2\theta\subrm{eff})$-space (lower panel) for selected fiducial sterile neutrino masses. Due to the higher non-linearity of the posterior distribution, the ensemble size has been increased to $N=10D$. Still, convergence is limited with a Gelman Rubin statistic $R<1.01$ only for $m_h=\SI{4}{eV}$ and $R<1.1$ for the other two examples. It can be seen that for $m_h = \SI{1}{eV}$ (left) neither the fiducial sterile neutrino mass nor the effective mixing angle can be reasonably estimated. Most plausibly, it is not possible to extract enough information from the integral $\upbeta$-spectrum if the active neutrino, sterile neutrino and right-handed current signatures are all together concentrated on a region scarcely larger than the energy resolution of $\approx \SI{1}{eV}$. For $m_h=\SI{2}{eV}$ (middle), the posterior distribution is significantly sharper. However, there is still a strong, slightly non-linear, correlation which leads to a rather large uncertainty on the effective left-right interference and effective mixing angle. The correlation pattern shows that it is still difficult to disentangle the signatures of the active-sterile mixing and the right handed currents, yet, there is a clear relation between both. For $m_h=\SI{4}{eV}$ (right), the uncertainties and correlations become smaller, allowing to define reasonable credible intervals for both the sterile neutrino and the right-handed current parameters at the same time. Still, the correlation between $c\subrm{LR}$ and $\sin^2\theta\subrm{eff}$ shows that for allowing $c\subrm{LR}$ to be a free parameter, one loses precision on estimating the mixing angle.

\section{Discussion}

It has been shown that KATRIN is sensitive to right-handed currents in combination with light sterile neutrinos. Without constrained endpoint, the average statistical sensitivity varies from about 0.5 to 0.05 in terms of $c\subrm{LR}$, depending on the sterile neutrino mass, plus a significant estimation bias. With a constrained endpoint, the sensitivity improves by up to an order of magnitude, depending on the prior uncertainty on the endpoint. 

\begin{figure}[hbt]
  \centering
  \includegraphics[width=\linewidth]{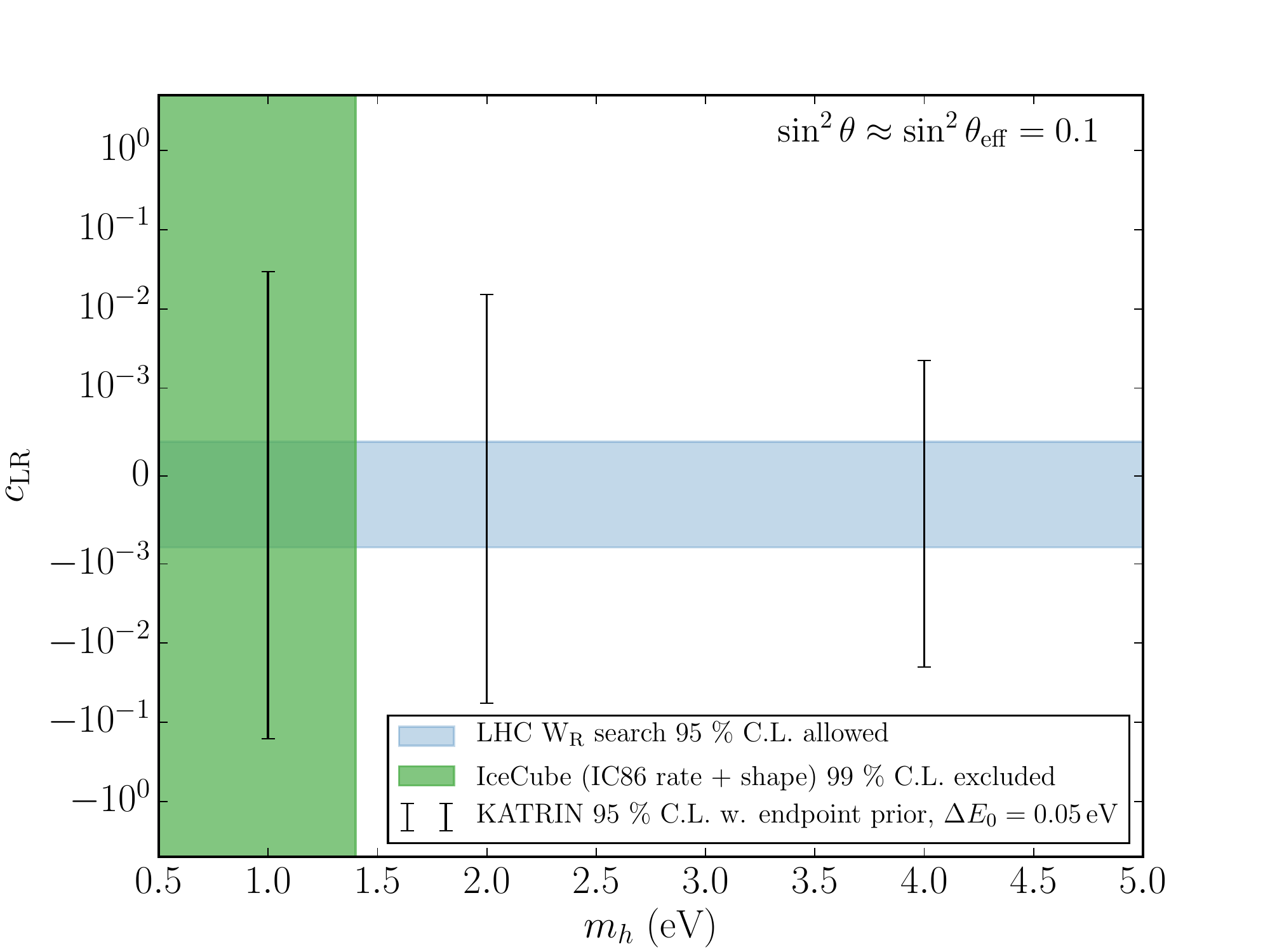}
  \caption{Bayesian credible intervals for configuration with endpoint prior \eqref{eq:endpointconstraint} of $\Delta E_0 = \SI{50}{meV}$ and $\sin^2\theta\subeff=0.1$ (black error bars, cf. fig. \ref{fig:e0pull-sens}), contrasted with 99 \% C.L. exclusion from IceCube \cite{IceCube2016} in the "IC86 rate+shape" analysis configuration for $\sin^2\theta = 0.1$ (green region) and limits from $W_R$ search at the LHC \cite{CMS2014, ATLAS2012} which state $m_{W_\up R} \gtrsim \SI{3}{TeV}$ at 95 \% C.L., translated into bounds on $c\subrm{LR}$ by eq. \eqref{eq:crh}, assuming $|\xi| \lesssim \num{e-3}$ \cite{Barry2013} for $\sin^2\theta = 0.1$ (shaded blue region).}
  \label{fig:sens-comparison}
\end{figure}

For a non-LRSM scenario of right-handed currents in absence of any sterile neutrino, it has been shown in \cite{Riis2011a} that KATRIN is unlikely to improve the limit, especially because of the correlation between endpoint and interference parameter. In the scenario with additional light sterile neutrinos, which has been investigated in the present work, KATRIN performs significantly better. If LRSM is assumed as underlying model, giving  rise both to sterile neutrinos and right-handed currents, the hard mathematical boundary $|c\subrm{LR}| < \sin^2\theta\subeff$ has to be kept in mind. The chances to significantly go below this hard limit rise with increasing mixing angle, increasing sterile neutrino mass and most importantly more stringent bounds on the endpoint. Additionally, it has been shown that, given a fit model with right-handed currents, the possibility of reasonably estimating the sterile neutrino mass and mixing angle is only given for higher masses $m_h \gtrsim \SI{2}{eV}$. That certainly is in conflict with the parameter space favored by the reactor neutrino anomaly \cite{LSND1998, MiniBooNE2007}. However, this parameter region has recently been excluded by IceCube \cite{IceCube2016}, which nevertheless still allows higher masses $m_h \gtrsim \SI{1.5}{eV}$ with at least $\sin^2\theta \lesssim 0.1$ (fig. \ref{fig:sens-comparison}).

Regarding the current experimental limits on a left-right symmetric $c\subrm{LR}$, the current LHC bounds roughly state $m_{W_\up R} \gtrsim \SI{3}{TeV} $ \cite{CMS2014, ATLAS2012} which can be translated via theoretical arguments into a bound on the LR mixing angle of about $|\xi| \lesssim \num{e-3}$ \cite{Barry2013}. The maximum left-right interference \eqref{eq:crh} is then given for a negative $\xi$ and vanishing $CP$ violating phase, which yields the bound $c\subrm{LR} \gtrsim - 0.003 \cdot \sin \theta$ for small $\theta$. On the other side, a bound of $c\subrm{LR} \lesssim 0.001 \cdot \sin \theta$ can be derived for $m_{W_\up R} \to \infty$ and $\xi > 0$ or $\cos\alpha <0$. While the KATRIN sensitivity is not able to surpass these boundaries (fig. \ref{fig:sens-comparison}), it nevertheless provides a useful complementary measurement without additional cost. Moreover, these bounds are only valid for LRSM-based right-handed currents, which require a right-handed weak boson. If a Fiertz-like interference term as in eq. \eqref{eq:beta-rh-eff} is caused by a different mechanism, there can well be a chance for KATRIN to test such models. 

As shown, the sensitivity and robustness on the method depends on the ability to constrain the endpoint. A significant improvement of the sensitivity and normalization of the estimation bias is only expected if the endpoint can be constrained with a precision better than \SI{0.1}{eV} (at $1\sigma$). The most promising way of achieving this aim are $^3$H-$^3$He mass measurements with precision Penning traps \cite{Myers2015,Streubel2014}. As future experiments aim for a bound of $\sim \SI{30}{meV}$ \cite{Streubel2014}, there is realistic hope to set stronger constraints on $E_0$ in the future. Since the endpoint is washed out by rotational-vibrational excitations of the daughter molecules, this means, regarding any future study of the systematics, that all molecular effects will need to be known sufficiently precise as well \cite{Bodine2015}. Also plasma-effects in the WGTS \cite{Kuckert2017} and the high-voltage stability \cite{Rest2017} will most likely have a non-neglegible effect. Since improvements of the bounds on the $Q$ value are expected in the near future, dedicated studies of how this can be translated quantitatively into a prior on $E_0$ are on the way. As KATRIN is expected to take first tritium measurements in the second half of 2017, the next milestone in the right-handed current search will be a test of the simulations on real data.

\acknowledgments

We would like to thank W. Rodejohann for discussions. This work is partly funded by BMBF under contract no. 05A11PM2, the IP@WWU program and DFG GRK 2149. 
\bibliography{library} 

\bibliographystyle{JHEP}

\end{document}